\documentclass[aps,prl,twocolumn,floats,showpacs,superscriptaddress]{revtex4}
\usepackage{graphicx,epsfig}% Include figure files
\usepackage{times}
\usepackage{graphics,dcolumn,bm,float}
\usepackage{amssymb,amsmath,rotate,color}

\usepackage{graphicx}

\begin{document}
\unitlength 1 cm
\newcommand{\be}{\begin{equation}}
\newcommand{\ee}{\end{equation}}
\newcommand{\bearr}{\begin{eqnarray}}
\newcommand{\eearr}{\end{eqnarray}}
\newcommand{\nn}{\nonumber}
\newcommand{\vk}{\vec k}
\newcommand{\vp}{\vec p}
\newcommand{\vq}{\vec q}
\newcommand{\vkp}{\vec {k'}}
\newcommand{\vpp}{\vec {p'}}
\newcommand{\vqp}{\vec {q'}}
\newcommand{\bk}{{\bf k}}
\newcommand{\bp}{{\bf p}}
\newcommand{\bq}{{\bf q}}
\newcommand{\br}{{\bf r}}
\newcommand{\bR}{{\bf R}}
\newcommand{\up}{\uparrow}
\newcommand{\down}{\downarrow}
\newcommand{\fns}{\footnotesize}
\newcommand{\ns}{\normalsize}
\newcommand{\cdag}{c^{\dagger}}

\title{Collective Spin and Charge Excitations in Planar Aromatic Molecules}

\author{K. Haghighi Mood}
\affiliation{Department of Physics, Science and Research Branch (IAU), Tehran, Iran}

\author{S. A. Jafari{\footnote {Electronic address: jafari@sharif.edu}}}
\affiliation{Department of Physics, Sharif University of Technology, Tehran 11155-9161, Iran}
\affiliation{School of Physics, Institute for Research in Fundamental Sciences, Tehran 19395-5531, Iran}
\author{E. Adibi}
\affiliation{Department of Physics, Isfahan University of Technology, Isfahan 84156, Iran}

\author{G. Baskaran}
\affiliation{Institute of Mathematical Sciences, Chennai 600113, India}

\author{M. R. Abolhassani}
\affiliation{Department of Physics, Science and Research Branch (IAU), Tehran, Iran}

\begin{abstract}
  Employing high accuracy fixed node diffusion Monte Carlo (DMC) method 
we calculated the lowest triplet collective excitation (spin gap), as well as an 
upper bound for the singlet excitations (charge gap) in a series of 
charge neutral planar non-ladder 
aromatic compounds. Both excitation energies lie below the continuum of particle-hole
excitation energies obtained from Hartree-Fock orbitals. Hence they can be interpreted
as genuine bound states in the particle-hole channel. 
Assuming a resonating valence bond (RVB) ground state which has been recently suggested
for $sp^2$ bonded systems 
[ M. Marchi, {\em et. al.}, Phys. Rev. Lett. {\bf 107}, 086807 (2011)],
offers a unified description of both excited states 
as two-spinon and doublon-holon bound states. 
We corroborate our interpretation, by  Exact diagonalization study of a 
minimal model on finite honeycomb clusters.
\end{abstract}
\maketitle

%\section{Introduction}
{\em Introduction}:
Correlation effects are characteristic of pi-$\pi$ conjugated systems composed essentially
of hexagonal arrangements of $sp^2$ bonds~\cite{Klein}. Pauling took the initiative to describe 
the bonding in benzene (C$_6$H$_6$), a prototype of these systems, in terms of valence 
bonds (VB), focusing his attention on the spin part of the bonding wave function, 
namely a singlet. Such singlet valence bonds can be formally described as the ground state
of an effective Heisenberg exchange interaction $J{\mathbf S}_1.{\mathbf S}_2$, where
$J$ is the exchange integral between the overlapping atomic orbitals~\cite{WhoIsRVB}. 
When the coordination number is low, in the above effective (model) Hamiltonian, 
the condition is more conducive for superposition of valence bond singlets 
to constitute the ground state -- a unique opportunity provided by three-fold 
coordination in these aromatic systems.
Hence Pauling's formulation of the energy levels of molecules in
terms of quantum mechanical superposition of valence bond configurations, the
so called resonating valence bond (RVB)~\cite{PaulingBook} 
becomes an important alternative rout to understand energy levels of 
molecules. 

Some recent works~\cite{Sorella,Azadi,HydrogenAdsorptionRVB}
have combined the power of Monte Carlo methods
with the basic notion of RVB~\cite{Anderson87}, 
to construct variational wave functions
in terms of geminals (determinants composed of "pairing" wave functions).
Optimization of such RVB based many-body wave-functions, determine the
properties of $sp^2$ bonded systems with remarkable 
accuracy~\cite{Sorella,Azadi,HydrogenAdsorptionRVB}. 
More specifically this technique captures the Kekule and Dewar contributions to the
ground state of benzene~\cite{Azadi}. Therefore, the
notion of RVB in these systems is capable of capturing interesting
many-body effects in the ground state, at much lower computational
cost, compared to more involved quantum chemical methods.
The application of the above method to undoped graphene indicates that
the ground state is a short range, gapped spin liquid~\cite{Azadi} which agrees
with other proposals based on the Hubbard model~\cite{Assaad, Vaezi, BaskaranJafari}.

In view of the above mentioned evidences for a possible spin liquid 
{\em ground state} in planar $sp^2$ bonded systems, 
the next natural question would be about the nature of excited states.
As a simple prototype which demonstrates the inadequacy of single-particle 
description of energy states in this family of molecules,
consider benzene, C$_6$H$_6$ for which the MO would predict a singlet ground state for 
six $p_z$ electrons on the hexagonal ring. Within MO picture, the first singlet excited 
state ($S_1$) and the first triplet excited state ($T_1$) are expected to be 
degenerate. However, observation of a remarkable splitting between the low-lying triplet and singlet 
excited states~\cite{daSilva} indicates the importance of correlation 
effects even in the excited states of these molecules.   
Since such correlation effects are based on local interactions,
one expects the same picture to hold even in extremely 
extended members of this family, such as graphene~\cite{NetoRMP}
and carbon nano-tubes. The weak coupling (itinerant) limit of the graphene 
is well known to represent a Dirac liquid~\cite{Jafari2009}, and can be 
described by standard single-particle approach~\cite{NetoRMP}. 
However, {\em ab-initio} calculations~\cite{Wehling} show that the strength of short-range 
part of the Coulomb interaction in these materials is $\sim 10$ eV, which is 
remarkably high and comparable to the estimated
values of these parameters in conjugated polymers~\cite{Baeriswyl}. 
For such large values of Hubbard parameter $U$ in these systems, emergence of a
non Fermi liquid state, such as spin liquid~\cite{Assaad,Mosadeq,Vaezi} becomes conceivable.

In this paper, we investigate the nature of low-lying excited states
in small molecules belonging to the family of $sp^2$-bonded carbon systems. 
Here we employ the state of the art QMC method to investigate
the nature of many-body excitations in such hydrocarbons.
This numerically accurate method suggests that the lowest excitation in such 
molecules is a triplet state, separated by a substantial
gap from the next singlet excited state, for which we obtain an upper bound. 
We argue that these two lowest excited states, namely, $T_1$ and $S_1$ can be naturally
understood in terms of a picture based on spin-charge separation.
This suggests that the ground state could be viewed as a resonating valence 
bond state, in agreement with a recent proposal by Marchi and coworkers ~\cite{Azadi}.

%\section{Method}
{\em Method}: 
Considering computational cost and accuracy, Variational Monte Carlo (VMC) 
and Diffusion Monte Carlo (DMC)~\cite{Foulkes} algorithms are methods of choice 
for the calculation of many body properties of medium electronic systems. 
These QMC methods can achieve chemical accuracy with a typical computational 
cost ranging from the second to fourth power of the number of 
particles~\cite{Needs}. In this paper we use these 
methods as implemented in CASINO package to calculate spin and charge gap of 
some aromatic compounds. 
The CASINO code employs important sampling DMC method~\cite{Umrigar,CASINOman}
to project out the many-body lowest energy state.
In this method, the important sampled imaginary time Schrodinger equation 
is of the following form:
\be
  f(\textbf{R},t+\Delta\tau)=\int K(\textbf{R},t+\Delta\tau\ ; \textbf{R}',t) 
  \ f(\textbf{R},t)\ d\textbf{R}',
   \label{imp.dmc.eqn}
\ee 
where $f({\textbf R},t+\Delta\tau\ ) =\Psi_t(\textbf{R}) \psi(\textbf{R},t+\Delta\tau\ )$, 
$\Psi_t(\textbf{R})$ is  the trial wave function and  $\psi(\textbf{R},t+\Delta\tau\ )$ 
is system wave function. The kernel $K(\textbf{R},t+\Delta\tau\ ; \textbf{R}',t)$ is 
the propagator. As $\Delta\tau$ approaches to infinity, 
$\psi(\textbf{R},t+\Delta\tau)$ tends to ground state in any sector 
corresponding to a definite set of quantum numbers.
For an efficient DMC calculation we need an optimized trial wave function. We used the
multiplication of spin up and down Slater determinants and a Jastrow factor as a trial wave function:
\be
  \Psi_t=\textit{e}^{\textit{j}(\textbf{R})}\textbf{\textit{D}}^{\upharpoonleft}(\textbf{r}_1,...,\textbf{r}_N)
  \textbf{\textit{D}}^{\downharpoonleft}(\textbf{r}_1,...,\textbf{r}_N).
  \label{jastrow-slater.eqn}
\ee 
Here $\textbf{R}=(\textbf{r}_1,\textbf{r}_2,...\textbf{r}_N)$ 
denotes the spatial coordinates of all the electrons. 
The single-particle orbitals employed in the above Slater determinants
have been constructed from  Hartree-Fock (HF) mean field solutions which serve
as a reference basis for "free" particle-hole excitations. Note that
this is not the exact Jastrow-Slater trial wave function form, as it is antisymmetric only 
with respect to the exchange of electrons with the same spin. Such wave functions can be used to obtain
expectation values with lower computational cost for any spin independent 
operators~\cite{Foulkes}. 
CASINO uses Jastrow factors of the form proposed in Refs.~\cite{CASINOman,Drummond}.
We have taken into account the electron-electron terms \textit{u},  
electron-nucleus terms $\chi$ centered on the nuclei and 3 body electron-electron-nucleus 
terms \textit{f} in our calculations:
\bearr
  j({r_{i}},{r_{j}}) =&& \sum_{i=1}^{N-1}\sum_{j=i+1}^{N} u(r_{i,j})
  +\sum_{I=1}^{N_{\rm ions}}\sum_{j=i+1}^{N}\chi_{I}(r_{i,I})\nn \\ 
  &&+\sum_{I=1}^{N_{\rm ions}}\sum_{i=1}^{N-1}\sum_{j=i+1}^{N} f_{I}(r_{i,I},r_{j,I},r_{i,j}).
  \label{jastrow.eqn}
\eearr 
Optimization with respect to the parameters contained in the Jastrow factor 
was achieved by a VMC variance minimization procedure~\cite{Foulkes}. 
After VMC optimization we used the so optimized wave function as a
DMC trial wave function. Optimization of Jastrow factors without optimizing orbitals did not affect 
the accuracy in our calculations. However, optimization of Jastrow factors provides a 
better trial wave function for DMC calculation by making it more efficient. 
At the final stage of calculation, DMC projects out the ground state from this trial wave function. 

Using the above method, we calculate the many-body ground state in a given 
sector corresponding to the conserved total $S_z$ and total number of particles $N$.
To extract information about spin-charge splitting from total energies,
we proceed as follows: 
Let $E_0(N_\up,N_\down)$ denote the ground state energy for a system
where $N_\sigma$ is the number of electrons, each carrying spin $\hbar\sigma/2$
with $\sigma=\pm$ corresponding to $\up$ and $\down$ spin orientations, respectively.
$N=\sum_{\sigma}N_\sigma=N_\up+N_\down$, as well as the total
spin component, $S_z=(N_\up-N_\down)/2$ are constants of motion and 
hence do not change the numerical projection by DMC procedure. 
Therefore quantum numbers $(N_\up,N_\down)$ appropriately label 
various sectors of the spectrum.

Let us define the  the spin gap ($\Delta_{\rm s}$) and charge gap ($\Delta_{\rm c}$) as,
\bearr
   \Delta_{\rm s}&\equiv& E_0(N_\up+1,N_\down-1)-E_0(N_\up,N_\down),
   \label{SpinGap.eqn},\\
   2\Delta_{\rm c} &\equiv\!& \left[E_0(N_\up\!+\!1,N_\down)\!+\!E_0(N_\up\!-\!1,N_\down)\right]\nn\\
   &&\!-\!2 E_0(N_\up,N_\down)
   \label{ChargeGap.eqn}
\eearr
where $(N_\up,N_\down)$ correspond to neutral system. In all compounds considered here, 
the total number of electrons, $N$, is even, so that the unpolarized configuration 
(i.e. the state with equal number of spin up and spin down electrons, $N_\up=N_\down$) 
turns out to be the ground state. The energy $E_0(N_\up,N_\down)$ 
of this state can be calculated as follows: We generate a trial 
wave function from HF method with fixed total charge (neutral) and one spin 
multiplicity. Now to calculate the spin gap, Eq.~\eqref{SpinGap.eqn}
we flip one of the spins from, e.g. $\down$ sector, without altering
the total charge. The lowest energy obtained by QMC procedure in this
sector will correspond to  $E_0(N_\up+1,N_\down-1)$. Note that in this
sector, the total charge is zero and spin multiplicity is three. 
Note that since the spin and spatial symmetries of the many-body
Hamiltonian are not broken by HF solutions, the corresponding symmetry attributes 
are not changed by QMC projection. This means that the energy of the $(N_\up+1,N_\down-1)$ 
state will represent any of the three degenerate states belonging to 
the triplet representation of the $SU(2)$ group.
Spin gap defined above, represents the exact value of triplet excitation energy.

Now let us discuss the physical meaning of the charge gap defined above:
Imagine an infinitely large system, with equal number of $\up$ and 
$\down$ spin electrons.  When an electron is moved from one point in the 
system to a distant location, the resulting excitation will be a 
doublon-holon pair. $\Delta_c$ is half of the average energy of a pair,
and hence can be interpreted as an upper bound for 
the energy of a single holon. 
To calculate the energy of such doublon-holon
pair, an approximate scheme is to isolate two small sub-systems surrounding 
the holon, and the doublon. In the absence of interactions,
the doublon-holon energy will be given by the first term in the right
hand side of Eq.~\eqref{ChargeGap.eqn}. However, in reality there will
be an attractive interaction between them which lowers their true
energy. Therefore $2\Delta_c$ defined
above, is an upper bound for the energy of doublon-holon pair with respect
to the neutral background. Because of the time-reversal symmetry of the
Hamiltonian employed here, for such excitations based on charge fluctuations
the spin orientation of the added/removed electron does not matter. 

 \begin{figure}[t]
  \begin{center}
   % \vspace{0.8 cm}
    \includegraphics[height=2.52cm]{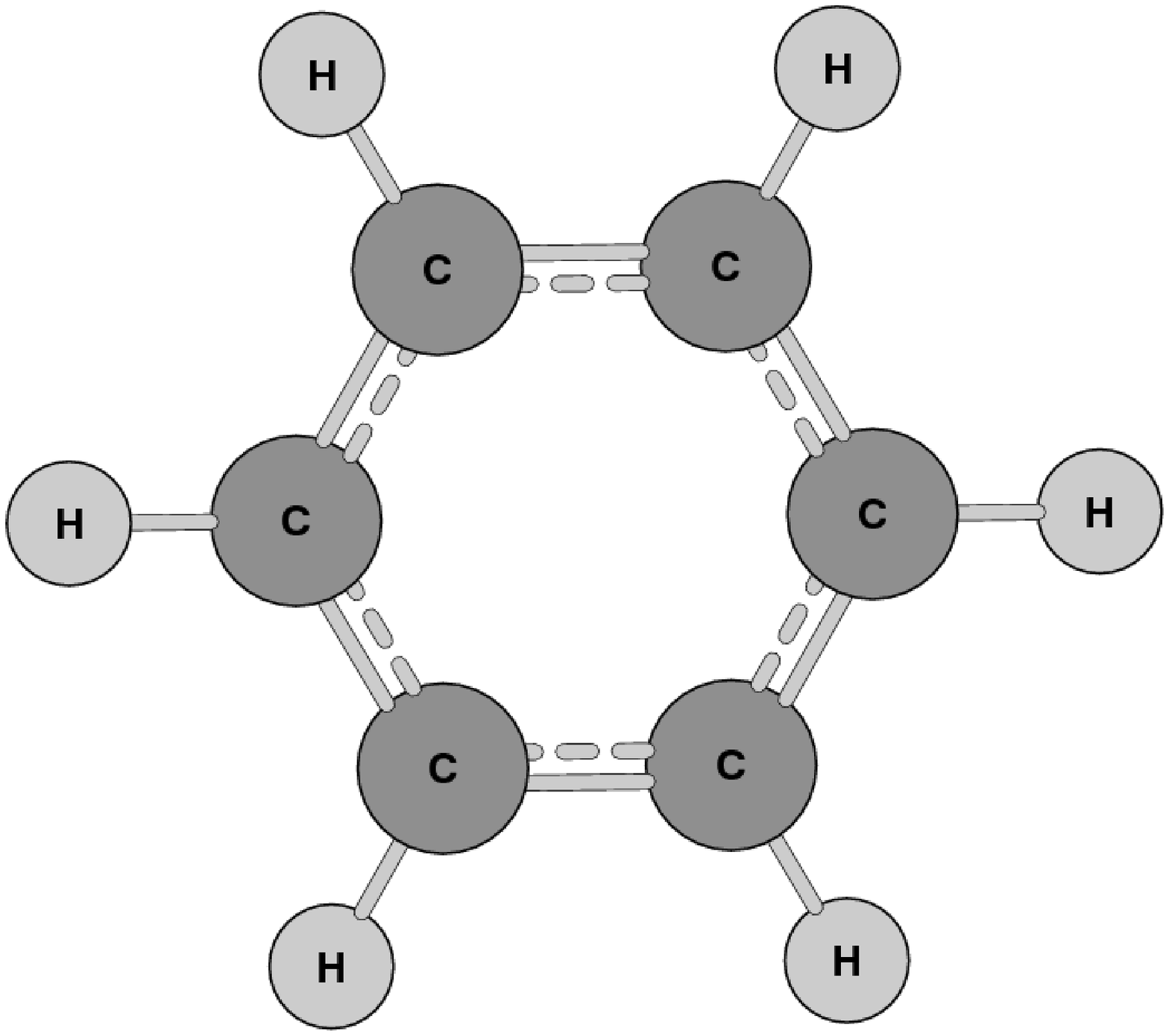}
    \includegraphics[height=2.52cm]{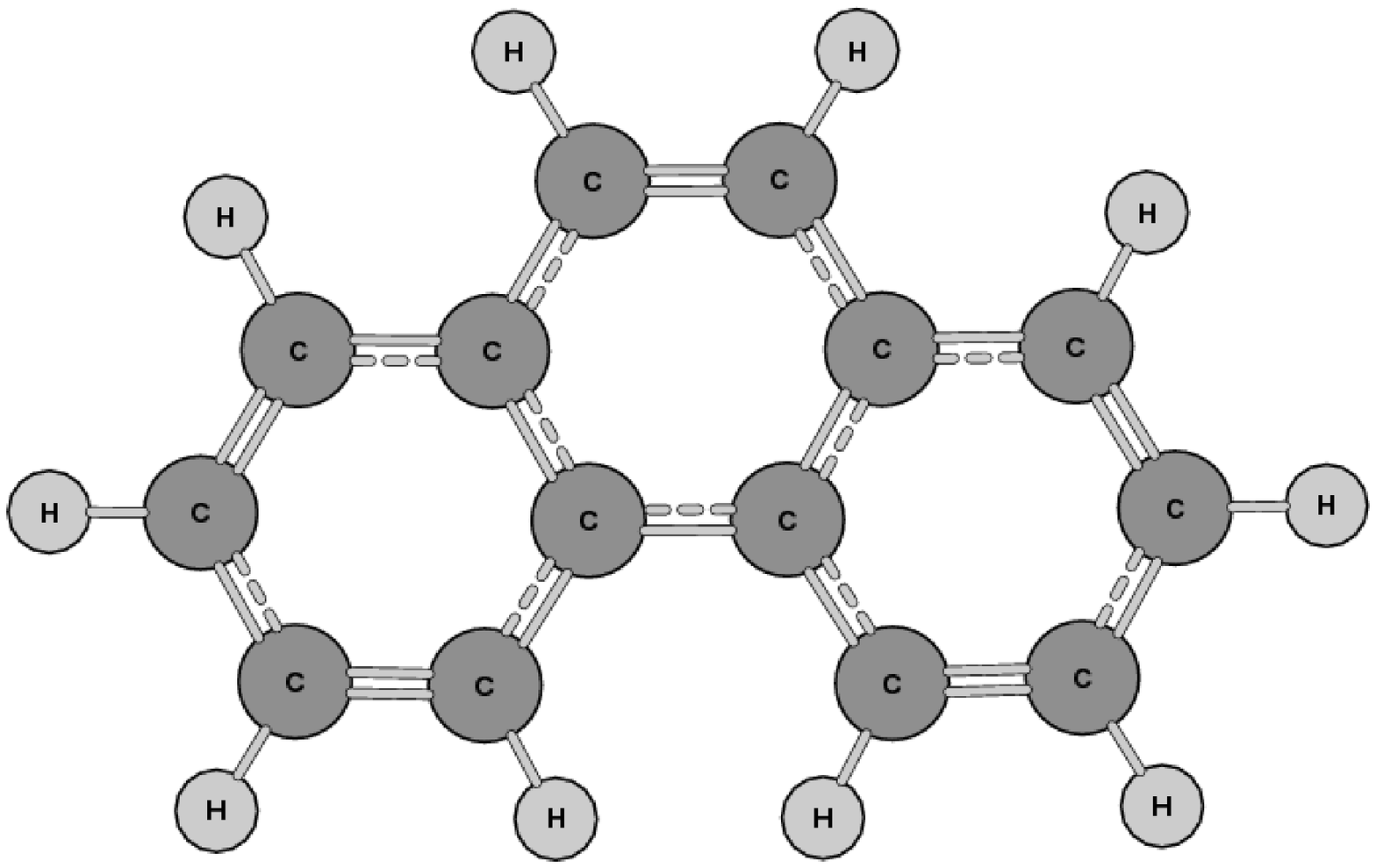}
    \includegraphics[height=3.0cm]{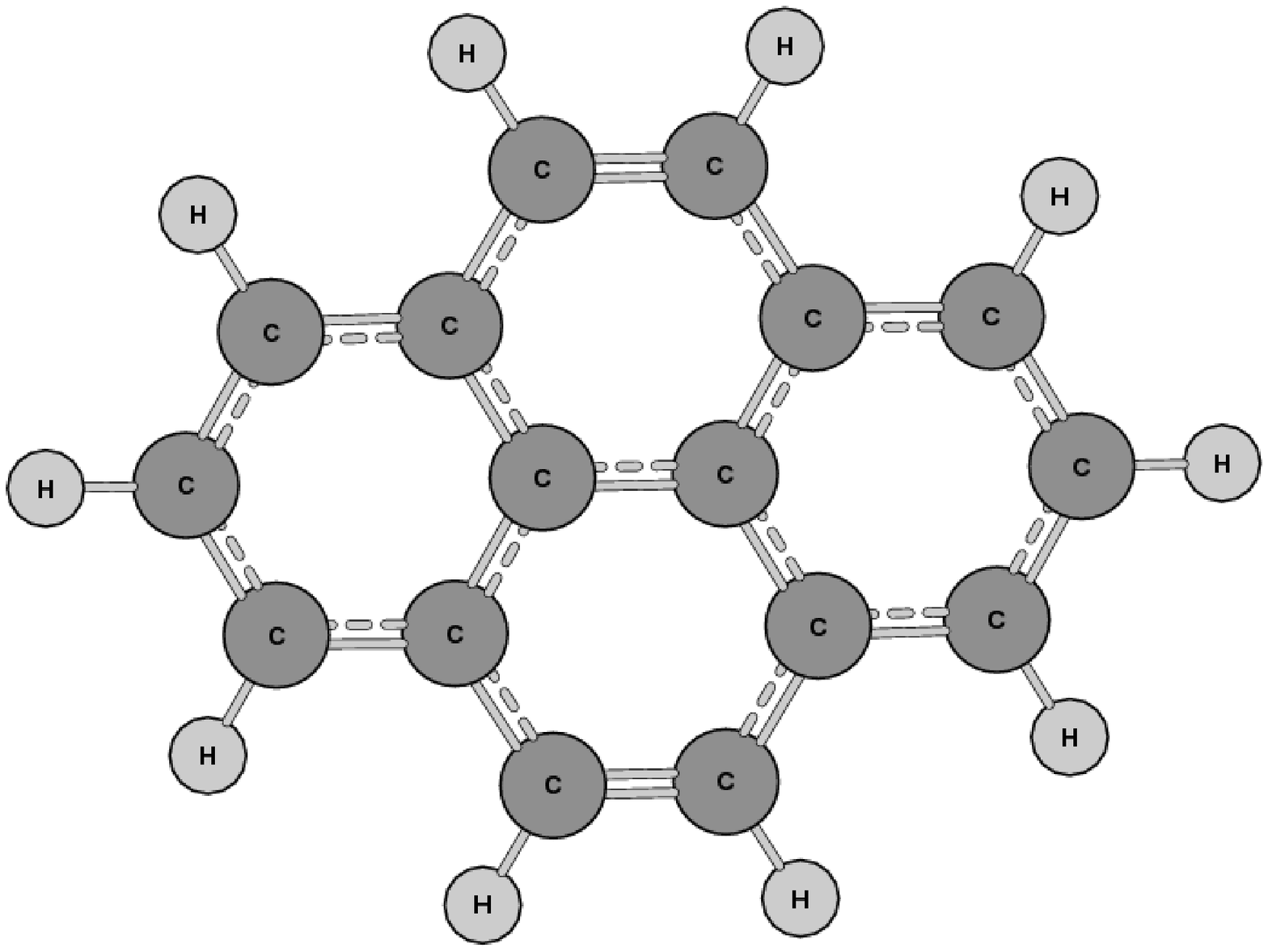}
    \includegraphics[height=3.0cm]{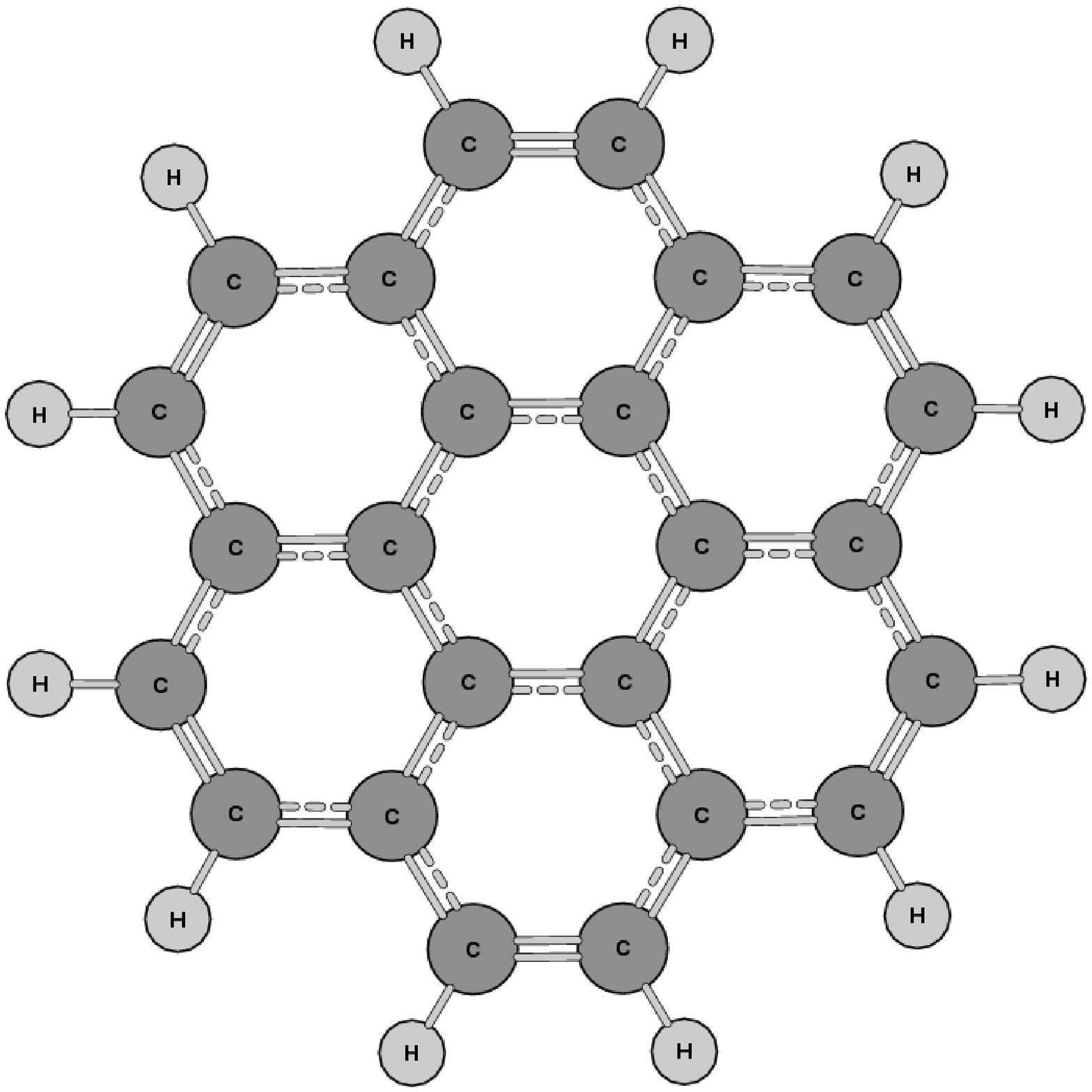}
    \caption{Benzene C$_6$H$_6$, Phenanthrene C$_{14}$H$_{10}$, 
    Pyrene C$_{16}$H$_{10}$, Coronene C$_{24}$H$_{12}$}
    \label{Benzene.fig}
  \end{center}
\end{figure}

\begin{table}[bt]
\caption{Spin and charge excitations (eV)}   % title of Table
\centering                          % used for centering table
\begin{tabular}{c c c c}            % centered columns (4 columns)
\hline\hline                        %inserts double horizontal lines
Compound & Spin Gap & $\Delta_c$ &  \\ [0.5ex]   % inserts table
%heading
\hline                              % inserts single horizontal line
C$_6$H$_6$ & $3.8(8)$ & $5.4(6)$  \\               % inserting body of the table
C$_{14}$H$_{10}$ & $3.4(1)$ & $4.3(5)$  \\
C$_{16}$H$_{10}$ & $2.4(9)$ & $3.8(2)$  \\
C$_{24}$H$_{12}$ & $3.0(6)$ & $3.6(7)$  \\
C$_{28}$H$_{14}$ & $2.7(8)$ & $3.4(0)$  \\ [1ex]         % [1ex] adds vertical space
\hline                              %inserts single line
\end{tabular}
\label{table:results.1}          % is used to refer this table in the text
\end{table}

\begin{figure}[t]
  \begin{center}
    %\vspace{0.8 cm}
    \includegraphics[width=7cm]{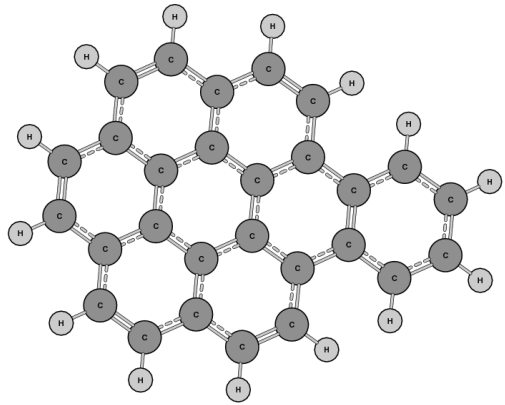}
    \caption{Coronene with additional benzene ring C$_{28}$H$_{14}$ (benzocoronene)}
    \label{Coronene.fig}
  \end{center}
\end{figure}

{\em Results:}
For five planar aromatic compounds depicted in Figs.~\ref{Benzene.fig},\ref{Coronene.fig}
we have calculated the above charge and spin gaps 
within the all electron fixed node DMC scheme. 
The results are reported in Table~\ref{table:results.1}. The spin gaps 
obtained here are in good agreement with experimentally reported values~\cite{daSilva}.
Also the charge gap $\Delta_c$ obtain here as an upper bound for the $S_1$ state
agrees with existing results. For example in case of benzene, $\Delta_c=5.4$ eV
represents a fair upper bound for the calculated result $E_{S_1}=4.9$ eV~\cite{daSilva}.
Therefore the method prescribed here to calculate the spin gap
does indeed give the "lowest" excited state, and also the doublon-holon
interpretation employed here does represent a true upper bound for the energy of $S_1$ state.
For geometry optimization as well as trial wave function 
generation, we used 6-311G** Gaussian basis which has been done by Gaussian 03 code~\cite{g03}.
Note that we performed separate optimizations for ground, and excited states.
All required energies are obtained with an accuracy better than $\sim 5$meV per atom.
For each of the compounds reported in Table~\ref{table:results.1},
and corresponding to each set of quantum numbers $(N_\up,N_\down)$, we
have optimized the geometry and the trial wave function constructed based 
on HF method. Then the Jastrow parameters have been optimized using variance
minimization VMC. All reported DMC results are all-electron calculations,
and we did not use any pseudo-potential. DMC time step is taken to be 
$0.002$ Hartree$^{-1}$. Optimized geometries have been verified to ensure
they do not contain imaginary frequencies. 

\begin{figure}[tb]
  \begin{center}
    %\vspace{0.8 cm}
    \includegraphics[width=8.5cm]{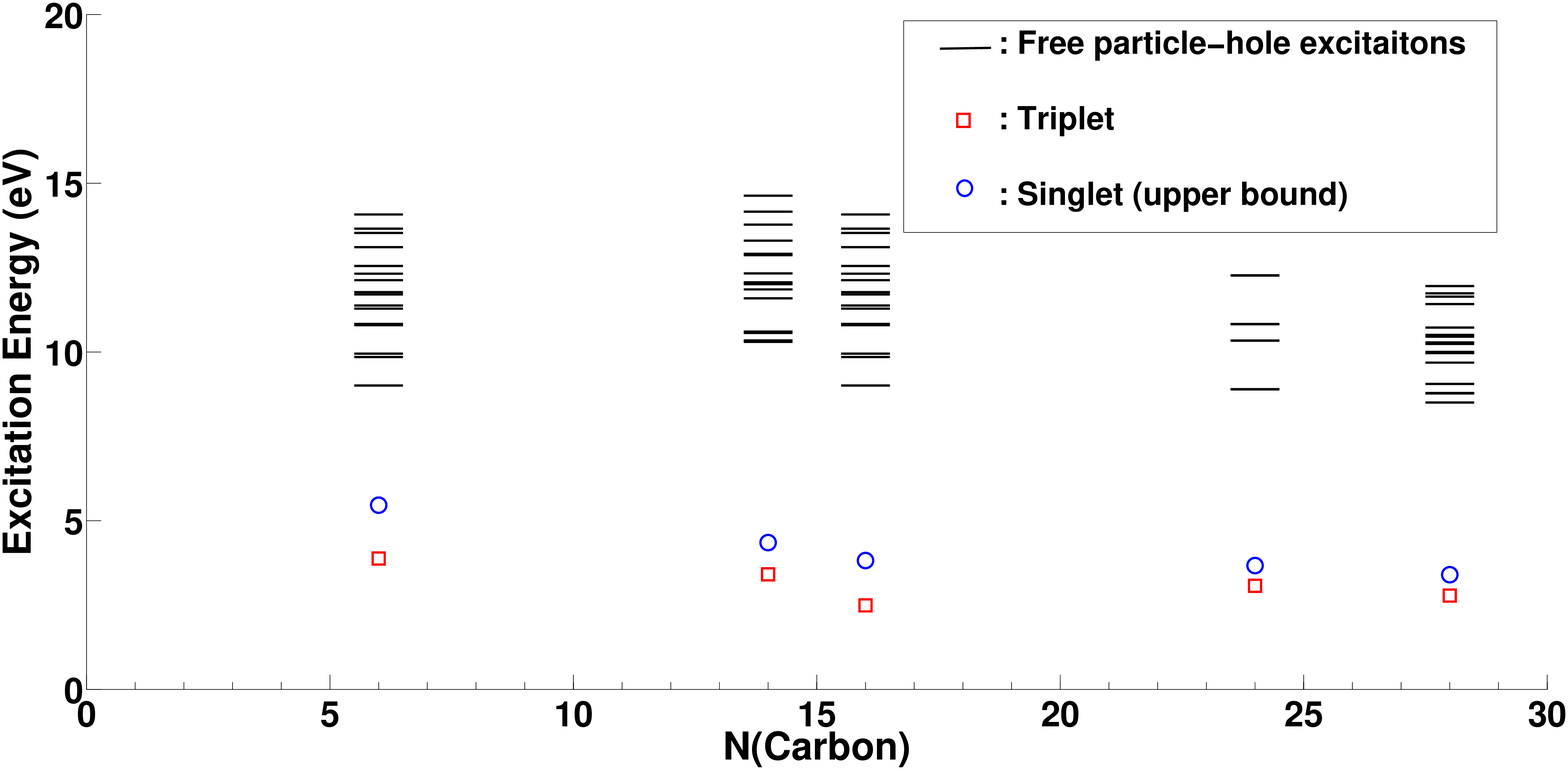}
    \caption{Charge and spin gaps versus the number of carbon atoms in aromatic compounds
    studied here. To generate the "free" particle-hole continuum, we have used the Hartree-Fock
    orbitals. For all studied compounds, energy of spin excitation, $T_1$ is below the 
    singlet (charge) excitation $S_1$, and they both are below the continuum of free
    particle-hole excitations.}
    \label{Gaps.fig}
  \end{center}
\end{figure}

To interpret the data in Table~\ref{table:results.1} let us represent them 
in a different way. 
In Fig.~\ref{Gaps.fig} we plot charge and spin gaps versus the number of carbons. 
The new physical interpretation come about, when we also plot the tower of particle-hole excitations
obtained from the (weakly correlated) Hartree-Fock theory for single particle states. 
This tower is the molecular analogue of the continuum of free particle-hole pairs. As can be 
seen by increasing the system size, the tower of particle-hole excitations approaches 
to a continuum. Moreover, the charge and spin gaps we obtain always remain below the 
the continuum of "free" particle-hole pairs. Therefore, they can be interpreted 
as the "bound state" of underlying free particle-hole pairs which are caused
by many-body effects. First important point which is suggested
by this figure is that, a very large energy difference between lower edge of the tower, 
and the many-body states found here implies they are long-lived excitations which
do not decay into the tower. Therefore, they can be associated with new
quasi-particles. Note that the blue circles $(\Delta_c)$ is an upper bound for
the true energy of a duoblon (holon), so that the true energy of the doublon 
state is even lower than $\Delta_c$. The question is, what are these quasi-particles?

Consider the lowest excited state $T_1$, which is a triplet many-body
state for all system sizes considered here. The $T_1$ state can be 
understood in terms of a simple RPA-like bound state formation in the
triplet channel of particle-hole pairs. A short range repulsion of 
Hubbard type translates into the attraction in the triplet particle-hole
channel, and binds them together~\cite{BaskaranJafari}. 
However the $S_1$ state whose exact location in Fig.~\ref{Gaps.fig}
is somewhere between the blue circle and red square can not
be understood in terms of simple RPA-like treatments, as the
RPA in singlet channel predicts an anti-bound-state {\em above}
the tower of free particle-hole states~\cite{BaskaranJafari}.
Therefore the second excited state $S_1$ is a genuine many-body effect, 
much beyond the simple RPA like treatments. The method used
here to obtain the upper bound for the singlet charge excitations
suggests that the $S_1$ can be associated to an average energy 
of a doublon and a holon. To corroborate this claim further, let us use
a simplified model Hamiltonian, which can capture the essence of the present
QMC calculation in a more transparent way. First of all note that the minimal model which
captures $T_1$ state is a Hubbard model. Moreover, our earlier 
study of the particle-hole excitation spectrum in 1D chains 
suggests that the singlet collective states below the particle-hole
continuum are controlled by the nearest neighbor Coulomb interaction~\cite{HafezJafari}.
Therefore the minimal effective model which captures both states
is an extended Hubbard model,
\be
   H=-t\sum_{\langle i,j\rangle \sigma} c^\dagger_{i\sigma} c_{j\sigma} +\mbox{h.c.}
   +U\sum_j n_{j\up}n_{j\down}+V\sum_{\langle\i,j\rangle} n_i n_j.
\ee
Here $i,j$ denote sites of a 2D honeycomb lattice and $\langle i,j \rangle$ 
indicates that they are nearest neighbors. $c^\dagger_{j\sigma}$ creates
an electron in the Wannier state corresponding to the $p_z$ orbital at
site $j$.  Here $U$ and $V$ denote the strength of on-site and nearest
neighbor Coulomb interactions. Estimates of these parameters based on the
{\em ab-initio} methods indicates that even the screened  of these 
parameters in graphene are substantial~\cite{Wehling} and on the
scale of $U\sim 10$ eV, which is comparable to corresponding estimates
for smaller aromatic molecules~\cite{Baeriswyl}. 
\begin{figure}[tb]
  \vspace{-0.5 cm}
  \begin{center}
    \includegraphics[width=8.0cm,angle=0]{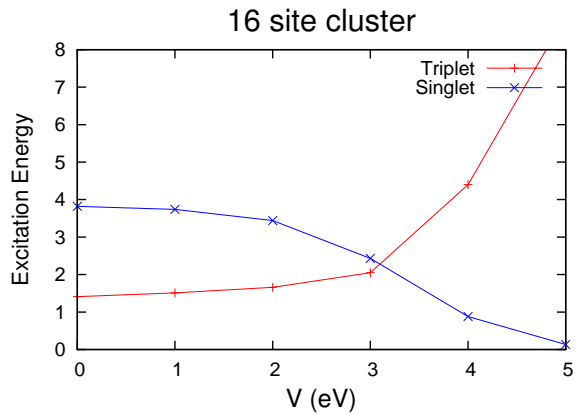}
    \caption{(Color online) The excitation energy for a $16$-site cluster corresponding to 
    C$_{16}$H$_{10}$ in the extended Hubbard model. Values of Hubbard $U$ and
    $t$ taken from Ref.~\cite{Wehling} are in eV. The ground state 
    always remains a singlet. The first excited states for small values of $V$
    are triplet. By increasing $V$, the order of $T_1$ (red) and $S_1$ (blue) is switched,
    and beyond $V\approx 3$ eV, $S_1$ will be the first excited state. 
    All energies in this figure are in eV.}
    \label{tUV.fig}
  \end{center}
\end{figure}

The result of the exact diagonalization for 
a $16$-site honeycomb lattice is shown in Fig.~\ref{tUV.fig}. 
The values of $U=9.3$ eV and $t=2.8$ eV are
adopted from Ref.~\cite{Wehling}. For the considered
range of $V$, the ground state always remains a total singlet state ($S_0$).
For small values of $V$, the first excited state is the $T_1$ triplet,
followed by a singlet excited state, $S_1$.
As $V$ increases, the singlet excited state, $S_1$ comes down and approaches
the energy of $T_1$ excited state for $V\approx 3$ eV. Beyond this point,
the first excited state will be a singlet state.
Thinking from the limit of very large molecules, the 
$S_1$ state will have no analogue in terms of plasmon oscillations.
Since, first of all, plasmon oscillations require non-neutral system~\cite{HwangSarma}
Secondly, long range Coulomb interaction 
makes the singlet branch either an acoustic plasmon (for 2D coulomb repulsion) 
or a gapped pi-plasmon (3D coulomb repulsion) branch. So the $S_1$ state
can not be interpreted as molecular analogue of plasmon mode. On the other hand, the decreasing
behavior of the $S_1$ energy with $V$ is consistent with
a doublon-holon interpretation: The repulsion $V$ among the electrons
will become attraction $-V$ between the doublon-holon pair, and
increasing $V$ will lower their energy.

{\em Summary and discussions}:
We have used {\em ab-initio} QMC method to obtain an accurate excited
state $T_1$ and an upper bound for the $T_1$. We then used exact diagonalization
to study a minimal model which captures the same set of excitations.
%Although $T_1$ can be understood within a simple RPA scheme, the $S_1$ state
%is consistent with an interpretation in terms of a doublon-holon formation.
Assuming RVB ground state~\cite{Azadi}, offers a unified understanding of both states.
In this scenario, the $T_1$ can be understood as the energy required to break a singlet
in the RVB background and render it triplet~\cite{Noorbakhsh}. Moreover, the $S_1$ 
can be attributed to a holon-doublon pair created by removing one electron from one carbon 
site, and placing it in the $p_z$ orbital of another carbon site. 
Such charge fluctuations are allowed because the on-site Coulomb energy $U$ is finite.
In this picture, the decrease in $S_1$ energy by increase in $V$ becomes quite
natural. 
%Since repulsion $V$, is translated to attraction between the
%holon and doublon, which prefers to create polar configurations 
%on a half-filled (undoped) background. When $V$ is very large, 
%these become softest excitations of an underlying many-body ground state.
This interpretation can be a possible
description of the collective charge excitations observed
in thick multi-wall carbon nano-tubes~\cite{Kramberger}.

%\section{Acknowledgement}
{\em Acknowledgements}:
K.H. thanks Mehdi D. Davari for useful discussions. 
S.A.J. was supported by the National Elite Foundation (NEF) of Iran.
We wish to thank Dr. M. Khazaei for assistance in the calculations,
and Dr. A. Vaezi and Prof. H. Fukuyama for useful discussions.

\end{document}